\documentclass[apj,onecolumn]{emulateapj}

\usepackage{amsmath}

\usepackage{epsf}

\renewcommand\email\texttt

\newcommand\ellg{{\ell_{\rm g}}}
\newcommand\bg{{b_{\rm g}}}

\newcommand\msun{{M_{\odot}}}
\newcommand\kms{{\rm kms}^{-1}}

\def\spose#1{\hbox to 0pt{#1\hss}}
\def\lta{\mathrel{\spose{\lower 3pt\hbox{$\sim$}}
    \raise 2.0pt\hbox{$<$}}}
\def\gta{\mathrel{\spose{\lower 3pt\hbox{$\sim$}}
    \raise 2.0pt\hbox{$>$}}}

\begin{document} 

\slugcomment{\sc submitted to \it the Astrophysical Journal}
\shorttitle{\sc An Orphan in the Field of Streams}
\shortauthors{\sc Belokurov et al.}

\title{An Orphan in the ``Field of Streams''} 
\author{V. Belokurov\altaffilmark{1},
N. W. Evans\altaffilmark{1}, 
M. J. Irwin\altaffilmark{1},  
D. Lynden-Bell\altaffilmark{1},
B. Yanny\altaffilmark{2},
S. Vidrih\altaffilmark{1}, 
G. Gilmore\altaffilmark{1}, 
G. Seabroke\altaffilmark{1},
D. B. Zucker\altaffilmark{1},
M. I. Wilkinson\altaffilmark{1},
P. C. Hewett\altaffilmark{1}, 
D. M. Bramich\altaffilmark{1}, 
M. Fellhauer\altaffilmark{1},
H. J. Newberg\altaffilmark{3},
R. F. G. Wyse\altaffilmark{4},
T. C. Beers\altaffilmark{5},
E. F. Bell\altaffilmark{6},
J. C. Barentine\altaffilmark{7},
J. Brinkmann\altaffilmark{7},
N. Cole\altaffilmark{3},
K. Pan\altaffilmark{7},
D. G. York\altaffilmark{8}}
\altaffiltext{1}{Institute of Astronomy, University of Cambridge,
Madingley Road, Cambridge CB3 0HA, UK;\email{vasily,nwe,mike@ast.cam.ac.uk}}
\altaffiltext{2}{Fermi National Accelerator Laboratory, P.O. Box 500,
Batavia, IL 60510}
\altaffiltext{3}{Rensselaer Polytechnic Institute, Troy, NY 12180}
\altaffiltext{4}{The Johns Hopkins University, 3701 San Martin Drive,
Baltimore, MD 21218}
\altaffiltext{5}{Department of Physics and Astronomy and Joint
Institute for Nuclear Astrophysics, Michigan State University, East Lansing, MI 48824}
\altaffiltext{6}{Max Planck Institute for Astronomy, K\"{o}nigstuhl
17, 69117 Heidelberg, Germany}
\altaffiltext{7}{Apache Point Observatory, P.O. Box 59, Sunspot, NM
88349}
\altaffiltext{8}{Department of Astronomy and Astrophysics, University
of Chicago, Chicago, IL 60637}

\begin{abstract}
We use Sloan Digital Sky Survey Data Release 5 photometry and
spectroscopy to study a tidal stream that extends over $\sim 50^\circ$
in the North Galactic Cap.  From the analysis of the path of the
stream and the colors and magnitudes of its stars, the stream is $\sim
20^{+7}_{-5}$ kpc away at its nearest detection (the celestial
equator). We detect a distance gradient -- the stream is farther away
from us at higher declination.  The contents of the stream are made up
from a predominantly old and metal-poor population that is similar to
the globular clusters M13 and M92. The integrated absolute magnitude
of the stream stars is estimated to be $M_r \sim -7.5$. There is
tentative evidence for a velocity signature, with the stream moving at
$\sim -40$ kms$^{-1}$ at low declinations and $\sim +100$ kms$^{-1}$
at high declinations. The stream lies on the same great circle as
Complex A, a roughly linear association of HI high velocity clouds
stretching over $\sim 30^\circ$ on the sky, and as Ursa Major II, a
recently discovered dwarf spheroidal galaxy.  Lying close to the same
great circle are a number of anomalous, young and metal-poor globular
clusters, including Palomar 1 and Ruprecht 106.
\end{abstract}


\section{Introduction}

The Sloan Digital Sky Survey (SDSS)~\citep{Yo00} is an imaging and
spectroscopic survey that has now mapped over $ 1/4$ of the sky. It
has already proven to be a powerful tool for the identification of
Galactic substructure. For example, \citet{Ne02} and \citet{Ya03}
identified a ring at low Galactic latitude, often called ``the
Monoceros Ring''. It spans over 100$^\circ$ in the sky~\citep[see
also][]{Ib03,Ro03}, but its progenitor remains unclear~\citep{Pe05}.
\citet{Od01} used SDSS data to find the spectacular 10$^\circ$ tidal
tails around the sparse and disrupting Galactic globular cluster Pal
5. More recently still, \citet{Be06a} found $4.5^\circ$ tails around
the high-latitude globular cluster NGC 5466, while \cite{Gr06a}
discovered a $63^\circ$ tail that presumably arises from a so far
unidentified globular cluster.

The best known example of a stream is the tidally stripped stars and
globular clusters associated with the Sagittarius dwarf spheroidal.  A
panorama of the the Sagittarius stream in the Northern hemisphere was
recently obtained by \cite{Be06b}, who mapped out the stars satisfying
$g-r < 0.4$ in almost all of SDSS Data Release 5 (DR5). This color
plot of the high-latitude Galactic northern hemisphere has been dubbed
the ``Field of Streams''. In addition to the features associated with
the Sagittarius dSph, the plot exhibits extensive substructure. The
purpose of this paper is to analyze the ``Orphan Stream'' -- so named
for its lack of obvious progenitor. This is a striking feature
discovered by~\citet{Be06b} in the Field of Streams, and independently
detected in public SDSS data by~\citet{Gr06b}.

\begin{figure*}[t]
\begin{center}
\includegraphics[height=8cm]{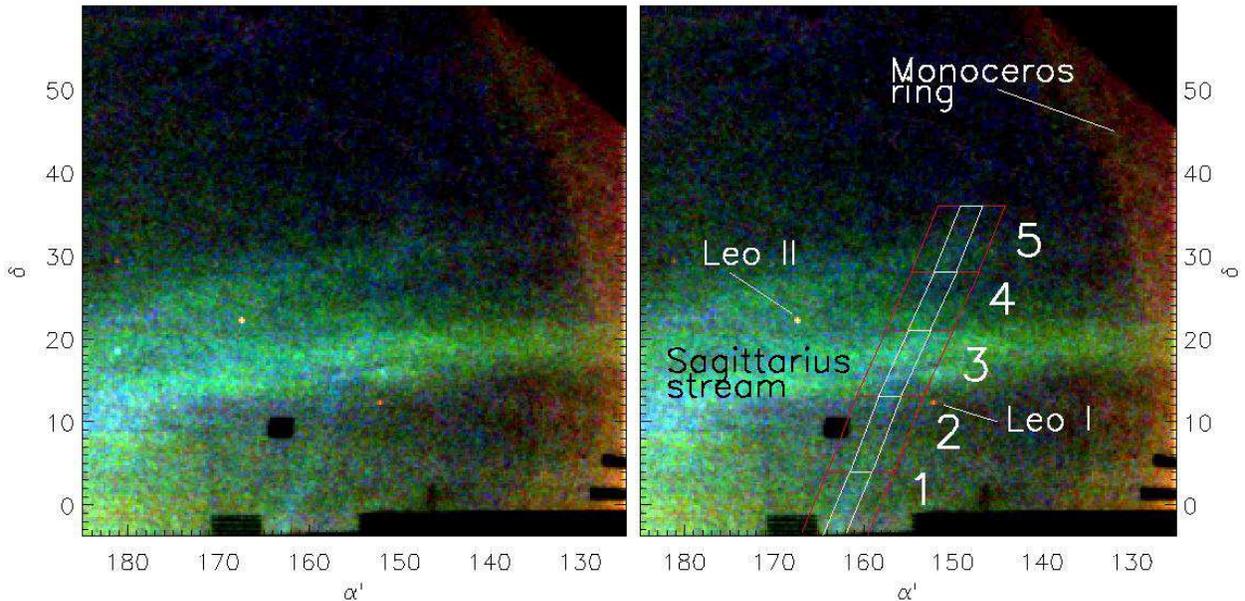}
\caption{\label{fig:vasa} Left: A false color RGB composite of density
of stars with $20.0 < r < 22.0$. Blue corresponds to $0.0<g-r \le0.2$, green
$0.2 < g-r \le 0.4$ and red $0.4 < g-r \le 0.6$. Right: The
Sagittarius and Monoceros structures are marked, together with the
on-stream and off-stream fields along the Orphan Stream. Also shown is
the location of the recently-discovered probable globular cluster
Segue 1~\citep{Be06c}. The black curves mark the limits of the DR5
spectroscopic footprint.}
\end{center}
\end{figure*}

\section{Morphology of the Orphan Stream}

SDSS imaging data are produced in five photometric bands, namely $u$,
$g$, $r$, $i$, and $z$~\citep[see e.g.,][]{Fu96,Ho01,Sm02,Gu06}. The
data are automatically processed through pipelines to measure
photometric and astrometric properties and to select targets for
spectroscopic follow-up~\citep{Lu99,St02,Pi03,Iv04,AM06}.  To correct
for Galactic reddenning, we use the maps of~\citet*{Sc98}. All the
magnitudes in the paper are reddenning corrected. Data Release 5
covers $\sim 8000$ square degrees around the Galactic North Pole, and
3 strips in the Galactic southern hemisphere.

The left panel of Figure~\ref{fig:vasa} shows the Orphan Stream in an
RGB composite image. It has been constructed using all SDSS DR5
stars~\footnote{A small number of Data Release 6 stars are used to
ensure continuous photometric coverage of the Orphan Stream
(c.f. Grillmair 2006b, who used Data Release 4 only).} with $20.0 < r
< 22.0$, with blue for stars with $0.0< g-r \le 0.2$, green for stars
with $0.2 < g-r \le 0.4$ and red for stars with $0.4 < g-r \le 0.6$.
The Orphan Stream is clearly visible running roughly from top to
bottom at right ascensions $\alpha \approx 150^\circ-165^\circ$. It
may be traced over nearly $50^\circ$ of arc in DR5. Some familiar
objects are marked in the right panel of Figure~\ref{fig:vasa},
including the bifurcated Sagittarius stream, the Monoceros Ring, the
newly discovered globular cluster Segue 1~\citep{Be06c} and the two
distant dwarf spheroidal galaxies Leo I and Leo II. On moving from
lower to higher declinations, the stream becomes less detectable.

\begin{figure}
\begin{center}
\includegraphics[width=15cm]{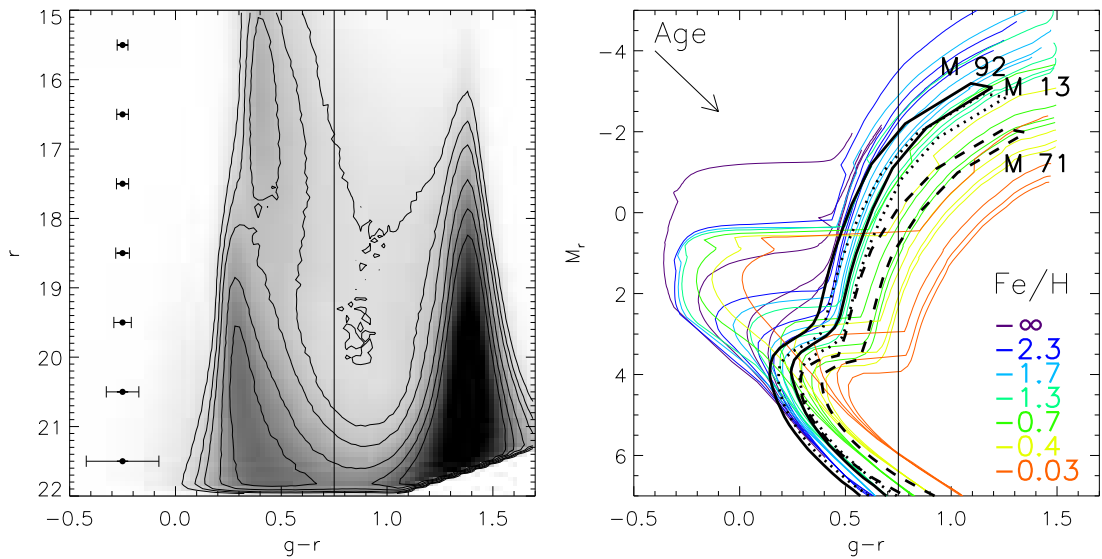}
\caption{\label{fig:hess} Left: Hess diagram for all stars in DR5. The
typical errors in color are shown as a column of error bars on the
left of the plot. The vertical line shows the color cut to constrain
the population to blue stars. Right: Isochrones from \citet{Gi04},
with different colors corresponding to different metallicities. Four
representative ages are shown for each metallicity (1, 5, 10 and 14
Gyrs, left to right).  Masks based on the ridgelines of M92 (solid),
M13 (dotted) and M71 (dashed) from \cite{Cl05} are also shown.}
\end{center}
\end{figure}
\begin{figure}
\begin{center}
\includegraphics[width=7cm]{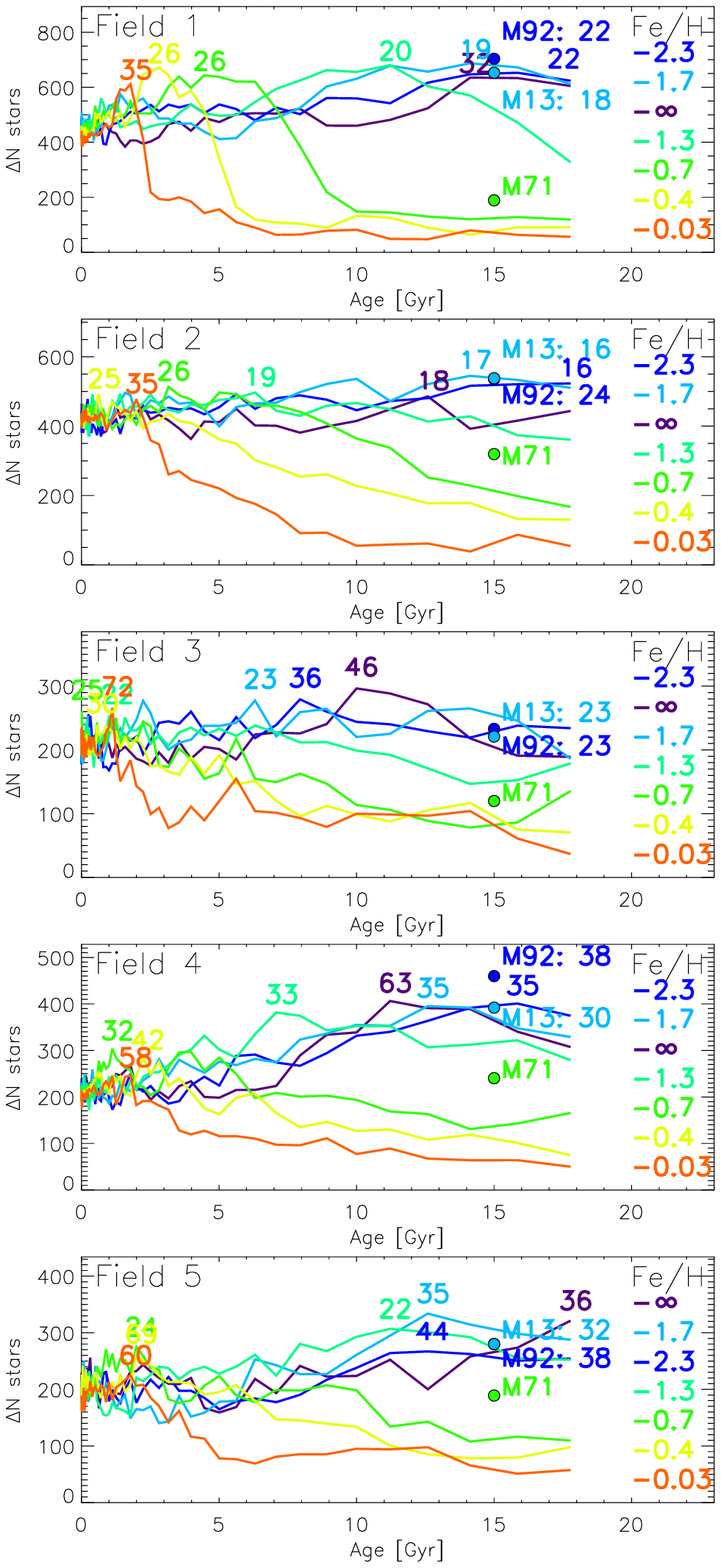}
\includegraphics[width=7cm]{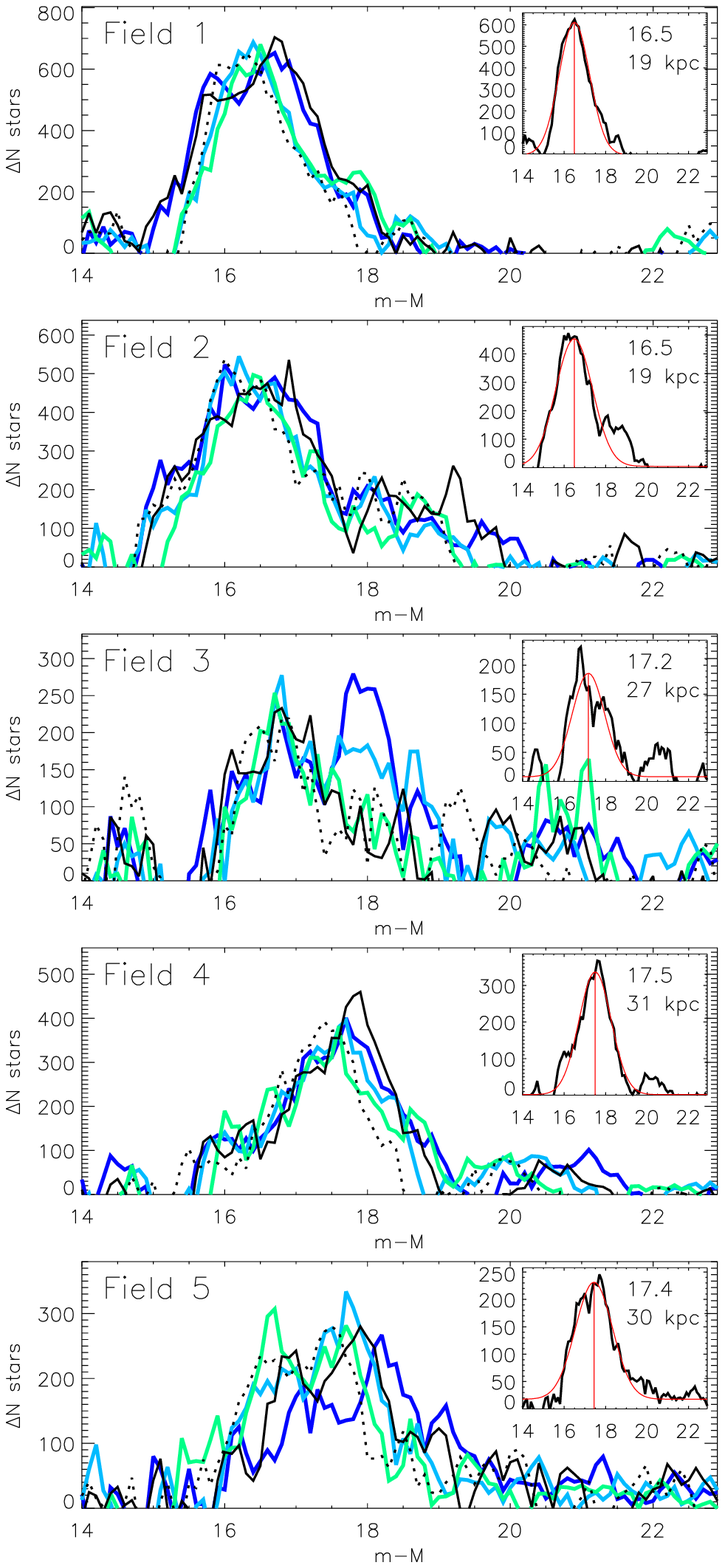}
\caption{\label{fig:stellarpops} Left: For each field, the excess of
stars in the Orphan Stream is shown as a function of age for different
metallicities. Each point on the curve gives the maximum number of
excess stars over all distance moduli. The peak of each curve gives
the maximal signal for a given metallicity and is marked with its
corresponding distance. It is apparent that there is a trade-off
between age, metallicity and distance.  Although there is a
degeneracy, old metal-poor populations perform slightly better than
young metal-rich ones. The filled circles show the signal picked up by
the ridgeline masks for the three clusters M92, M13 and M71.  Right:
For three metal-poor populations ([Fe/H] $= -2.3, -1.7, -1.3$), we
show the excess stars as a function of distance modulus for the age at
the peak of the curve in the left panel. Also shown as solid and
dotted lines are the curves using the ridgelines for M92 and M13.  The
inset shows the average of the five curves, together with a Gaussian
fit in red. The number gives the central value of the Gaussian, which
is our best estimate for the distance modulus, and hence the distance
in kpc.}
\end{center}
\end{figure}

\section{Distances and Stellar Populations of the Orphan Stream}

Our aim is to constrain the distance and the stellar content of the
Orphan Stream by comparison with CMDs of populations of known age and
metallicity.  For this purpose, we create masks (or color-magnitude
boxes) based on the $r$ versus $g\!-\!r$ ridgeline of globular
clusters and theoretical isochrones by shifting them both horizontally
and vertically.  The mask is applied to select stars from the fields
on and off the stream shown in the right panel of
Figure~\ref{fig:vasa}.  The signal is assessed by measuring the
difference in the number of stars between the on and off streams as a
function of distance modulus.

The left panel of Figure~\ref{fig:hess} shows a Hess diagram using all
the stars in DR5. From blue to red, the main components of this CMD
are the halo, thick disk and thin disk of the Milky Way.  The vertical
line shows a color cut at $g-r = 0.75$ used to minimise contamination
from the disc stars.  The right panel shows a sequence of theoretical
isochrones computed by ~\citet{Gi04} for the SDSS photometric system.
Different metallicities are distinguished by different colors as shown
in the key to the panel. Although we investigate a range of ages, only
the isochrones corresponding to 1, 5, 10 and 14 Gyrs are displayed.
To compare the model predictions with observations, we also use the
illustrated masks based on the ridgelines of the old (14 Gyrs)
clusters M92 ([Fe/H] = $-2.2$), M13 ([Fe/H] = $-1.6$) and M71 ([Fe/H]
= $-0.7$). These are produced from the data of \citet{Cl05} and chosen
to span a representative range of metallicities.

The Orphan Stream (and parts of the Sagittarius stream) shown in
Figure~\ref{fig:vasa} are blue-green in our RGB scheme. This
corresponds to $g-r \sim 0.3$, which is typical for the old, metal-poor
halo. Most of the stars are redwards of $g-r \sim 0.2$. The stars that
are bluewards could be either blue stragglers/blue horizontal branch
stars, or main sequence turn-off stars scattered by large photometric
errors.

It is well-known that, given a color-magnitude diagram, simultaneously
fitting for age, metallicity and distance modulus is
degenerate. However, we have some clues as to the likely nature of the
solution. The length and the width of the Orphan Stream suggests that
it is dynamically old. If star formation ceased after the interaction
that produced the tails, then the stars in the Stream are expected to
be old.  Also, when viewed from the Sun, tidal streams deviate from a
great circle. The deviation is controlled by $\langle D \rangle
/R_{\rm GC}$, where $\langle D \rangle$ is the stream's average
heliocentric distance and $R_{\rm GC}$ is the offset of the Sun from
the Galactic Center (here taken as 8 kpc). This parallactic effect
gives an average heliocentric distance to the Stream of $\sim 15 \pm
5$ kpc. 

The results of applying the masks to the photometric data are shown in
Figure~\ref{fig:stellarpops}. In the left column, each panel
corresponds to a different field. For a given metallicity and age, the
masks are applied at different distance modulus and the number of
excess stars in the on-stream field is calculated. The maximum value
fixes a distance modulus. The number of excess stars at this distance
modulus is used to build up the curve. The different colored lines
correspond to different metallicities.  Even though there is a
degeneracy, the distances to which models of different metallicities
converge are broadly consistent. For example, in Field 1, most of the
distances are between 20 and 30 kpc, whereas in Field 5, they are
mostly above 30 kpc.  These values are somewhat higher than the $15
\pm5$ kpc obtained from the parallactic effect, but consistent given
the uncertainties and given the strong assumptions in the latter
calculation.  The M92 and M13 masks perform similarly to the
corresponding isochrones, although there are some small discrepancies.
It is reassuring that the results based on theoretical models are
supported by those based on data.  

Even though young stellar population seemingly perform quite well in
Figure~\ref{fig:stellarpops}, this is not really the case. From the
turn-off color of the Orphan Stream (see e.g., Figure~\ref{fig:vasa}),
we can exclude young populations with a turn-off bluer than $g-r \sim
0.2$. Young isochrones are so blue that even the metal-rich ones
extend blue enough to overlap with the Orphan Stream CMD and hence
give a signal in Figure~\ref{fig:stellarpops}.

The general conclusion is that the old, metal-poor masks perform
better, and in some instances, significantly better than the
metal-rich masks.  Given this, we take the three metal-poor models,
namely [Fe/H] $= -2.2, -1.7, -1.3$, and identify the age at which the
signal is maximised. These masks are now used to investigate the
behavior of the signal as a function of distance modulus, shown in the
panels in the right column. Also shown are the curves based on the
ridgelines of M92 (solid) and M13 (dotted). They all show a similar
performance and so we average them as displayed in the inset (solid
black curve). A Gaussian model is then fit to this curve to give a
central value, its uncertainty and a dispersion. These numbers are
recorded in Table~\ref{tab:orphan}. The uncertainty in the mean gives
a lower bound to the true distance error, whilst the dispersion
overestimates the error, as the curves in the insets are a convolution
of the true error distribution with the mask.

\citet{Gr06b} has also recently analyzed the stars of the Orphan
Stream.  Grillmair's detection method uses M13 as a template for the
stellar population of the stream. His Figure 3 shows a clear upper
main sequence and sub-giant branch -- which appear to be located
somewhat blueward of the M13 ridgeline.  This is consistent with a
stream composed of a somewhat younger and/or metal-poor stellar
population than M13. In turn, this is in agreement with the blue color
of the Orphan Stream in our Figure~\ref{fig:vasa}, suggesting that our
detection method is picking up primarily turn-off and upper main
sequence stars.

\begin{table}
  \centering
  \caption{Positions, distance moduli and distances of the Orphan Stream.}  
  \label{tab:orphan}
  \begin{tabular}{@{}ccccc}
    Field & $\alpha$ & $\delta$ & $m-M$ & $D$ \\ \hline
    1 & $162.1^\circ$ & $-0.5^\circ$ & $(16.5 \pm 0.1) \pm 0.7$ & $(20
    \pm 1)^{+7}_{-5}$ kpc \\
    2 & $158.9^\circ$ & $8.5^\circ$  & $(16.5 \pm 0.1) \pm 0.9$ &
    $(20\pm 1)^{+10}_{-7}$ kpc \\
    3 & $155.4^\circ$ & $17.0^\circ$ & $(17.1 \pm 0.1) \pm 0.7$ &
    $(26\pm 1)^{+10}_{-7}$ kpc \\
    4 & $152.3^\circ$ & $25.0^\circ$ & $(17.5 \pm 0.1) \pm 0.8$ &
    $(32\pm 1)^{+13}_{-10}$ kpc \\
    5 & $149.4^\circ$ & $32.0^\circ$ & $(17.5 \pm 0.1) \pm 0.9$ &
    $(32\pm 1)^{+15}_{-12}$ kpc \\
    \hline
  \end{tabular}
\end{table}
\begin{figure}
\begin{center}
\includegraphics[width=10cm]{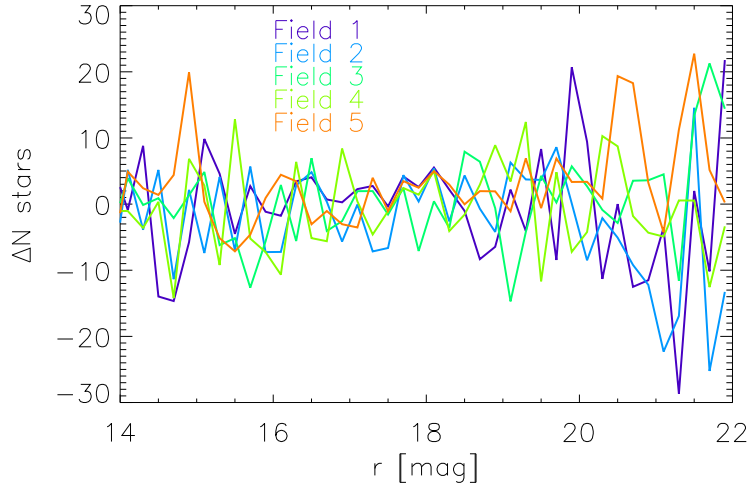}
\caption{\label{fig:bhbs} Differential histograms of number of stars
with $g-r < 0.1$ as a function of apparent magnitude in different
fields.}
\end{center}
\end{figure}

Figure~\ref{fig:bhbs} shows the results of a search for blue
horizontal branch (BHB) stars in the Orphan Stream. These are
identified with the cut $g-r < 0.1$, and as usual the difference
between on-stream and off-stream fields is computed. BHB stars have an
absolute magnitude of $M_r = 0.75$ (see the right panel of
Figure~\ref{fig:hess}). Such a BHB population would be detected by
peaks in the differential number distribution in the apparent
magnitude range $16< r <18$ for the five fields inspected. There is no
evidence for such a signal. However, blue stragglers have an absolute
magnitude $M_r$ of between 2 and 4 and there are hints of peaks in
fields 1 and 5 for $r \approx 20$. This seems consistent, given the
distance moduli in Table~\ref{tab:orphan}.

\section{Absolute Magnitude and Width of the Orphan Stream}

\begin{figure}
\begin{center}
\includegraphics[height=8cm]{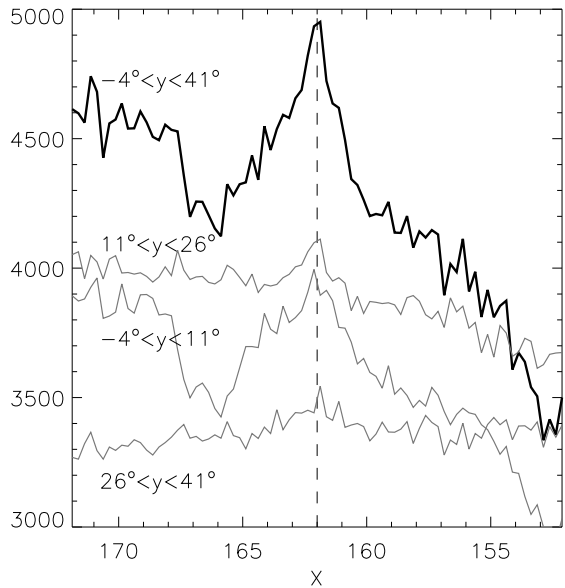}
\caption{\label{fig:dist} Profiles in stars with $g\!-\!r< 0.4$ and
$21.0 < r <22.0$ across the stream in a rotated coordinate system in which
the stream lies along the $y$-axis. Notice that the peak of the
profile barely shifts as we move along the stream. }
\end{center}
\end{figure}

Figure~\ref{fig:dist} shows cross-sections across the stream in a
coordinate system which has been rotated so the stream lies along the
$y$-axis. The cross-section contains only those stars with $g\!-\!r<
0.4$ and $21.0 < r <22.0$. We estimate that the thickness (FWHM) of
the Orphan Stream is $\sim 2^\circ$ in projection (or $\gta 650$ pc
assuming a conservative distance of $\sim 20$ kpc), which would make
it broader than all known globular cluster streams, but smaller than
the branches of the Sagittarius stream and the Monoceros Ring. This
suggests that the progenitor was a low luminosity dwarf satellite
galaxy, rather than a globular cluster.

Taking the total profile (upper curve in Figure~\ref{fig:dist}), we
fit a polynomial to estimate the background and then compute the
luminosity of the stream. The number of excess stars in the $45^\circ$
arc of the stream is $\sim 4110$ within $160^\circ < x < 164^\circ$.
Hence, the total apparent magnitude of the stream in these stars is
around $r \sim 12$ (assuming an average $r$ magnitude of 21.0).  Given
the stream FWHM of $\sim 2^\circ$, the average density of these stars
is $\sim 45.6$ per square degree, which translates to an average
surface brightness of the stream of $\sim 34\fm6$ per square
arcsec. To correct for other stars, we use the data on the luminosity
function of M92 derived from CFHT observations (Clem 2006, private
communication).  This gives the average surface brightness of the
stream as $\sim 32\fm4$ per square arcsec and the total $r$ magnitude
as $\sim 9.8$.  Assuming the distance really is $\sim 20$ kpc gives an
absolute magnitude of the $45^\circ$ arc as $M_r \sim -6.7$ Assuming
plausibly that there ought to be at least the same again on the other
side of the progenitor, then we can boost the total to at least $M_r
\sim -7.5$ for the stream stars alone. This number must be augmented
by the (unknown) contribution from the progenitor nucleus, if it is
still intact.

\begin{figure}
\begin{center}
\includegraphics[height=8cm]{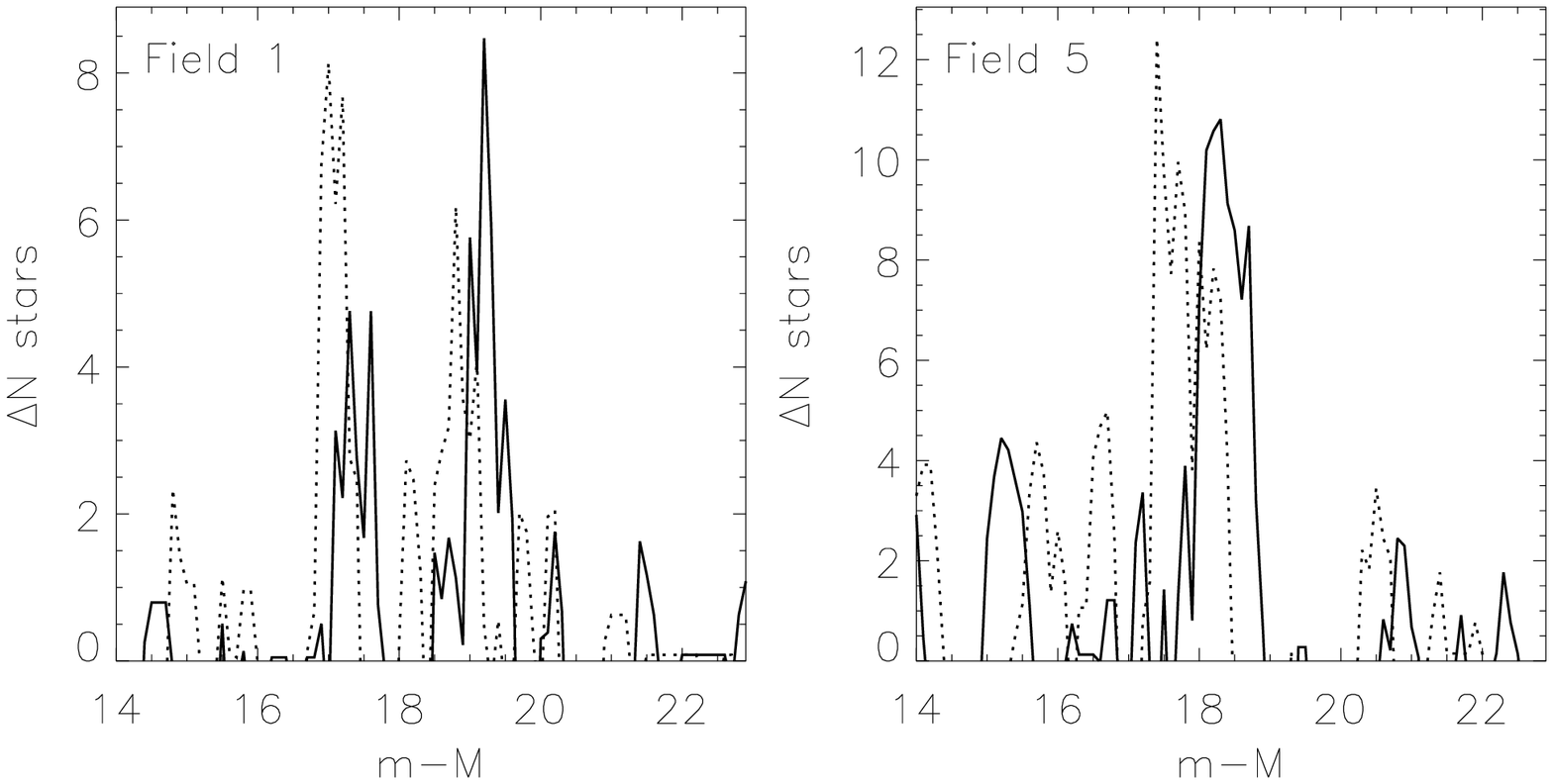}
\includegraphics[height=8cm]{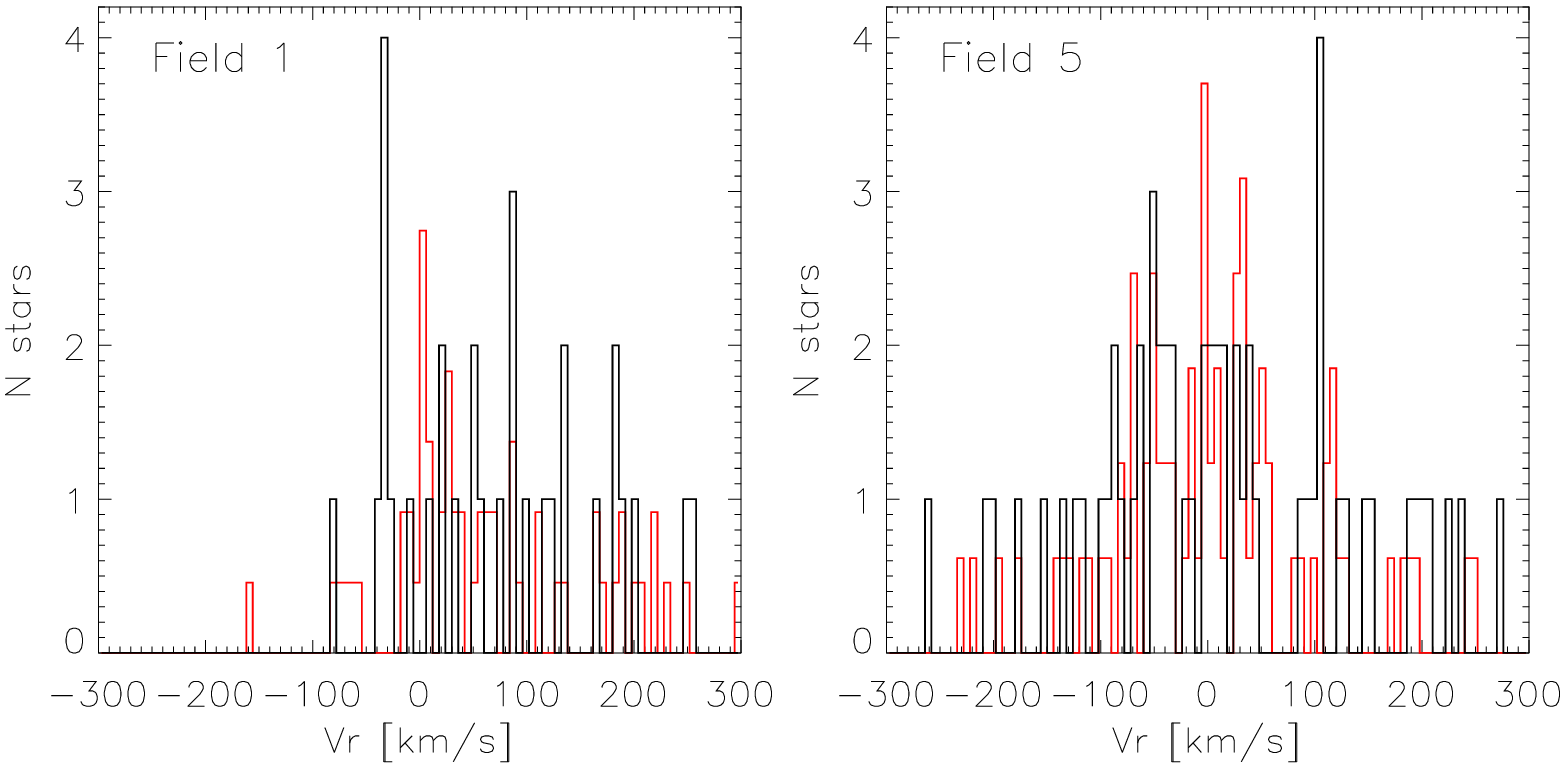}
\caption{\label{fig:velocities} Upper: Excess number of stars in the
on- and off-stream fields in the DR5 spectroscopic catalogue using the
M92 (solid) and M13 (dotted) ridgeline masks. Even though the
catalogue is much smaller, there is a reasonably clear detection of
the Orphan Stream at a distance modulus of $\sim 17$. Lower: Velocity
histograms of stars selected with the M92 and M13 ridgeline masks at
the distance moduli satisfying $16.5 \le m-M \le 17.5$ in Field 1 and
$17.5 \le m-M \le 18.5$ in Field 5. Black represents on-stream and red
off-stream field stars.}
\end{center}
\end{figure}
\newpage

\section{Velocities of the Orphan Stream}

The spectroscopic portion of SDSS DR5 covers $5713$ deg$^{2}$
~\citep{Ad06}. The outline of the spectroscopic footprint is shown as a
black line in Figure~\ref{fig:vasa}. It does not cover Fields 3 and 4
of the Orphan Stream. The number of stars in the spectroscopic database
is $\sim 215\,000$.  Only stars brighter than $g \sim 20$ were
targeted, and the selection algorithm is strongly non-uniform. A
large number of spectra of standard stars were taken for calibration
purposes. Other targets include K giants and dwarfs, F turn-off and
subdwarf stars, as well as BHBs.

The accuracy of the radial velocities measured from the $R = 2000$
spectra obtained by the SDSS spectrographs vary from 7 km/s for the
brighter stars ($r \sim 14$) to on the order of 20 km/s for the
fainter stars ($r \sim 20$)~\citep{Al06}, based on empirical
tests. Information on the stellar atmospheric parameters (T$_{\rm
eff}$, $\log$ g , [Fe/H]) for the DR5 stars with adequate spectra are
presently being obtained by application of an automated spectroscopy
analysis pipeline~\citep{Bee06}. This information will be utilized in
future discussions of the Orphan Stream, once a comparison of the
pipeline-derived atmospheric parameters with independently obtained
high-resolution spectroscopy has been completed~\citep{Si05}.

To test whether there is a signal of the Orphan Stream in the
spectroscopic database, we apply the masks of M92 and M13 discussed
earlier to the on- and off-stream fields. Given the faint magnitude
cut-off, there are no turn-off stars belonging to the Stream in the
database. So, our earlier algorithm is slightly changed to include a
color cut designed to pick up giant branch stars. It is reassuring to
see the peaks in the numbers of excess stars in the upper panel of
Figure~\ref{fig:velocities} at similar distance moduli to those
reported in Table~\ref{tab:orphan} for Fields 1 and 5 (no signal was
detected in Field 2).

We now select stars using the masks of M92 and M13 placed at the
distance moduli indicated by the peaks in the upper panels. Histograms
of the velocities of these stars for the on and off stream fields are
shown in black and red in the lower panels. A possible signature of
the Orphan Stream is a peak in black without a corresponding peak in
red. The largest peak in Field 1 is at $\sim -40$ kms$^{-1}$ and there
is no signal in red. The largest peak in Field 5 is at $\sim 100$
kms$^{-1}$ and similarly there is a low signal in red. Bearing in mind
the sparsity of the data, these detections are suggestive rather than
conclusive.

\begin{figure*}[t]
\begin{center}
\includegraphics[height=5cm]{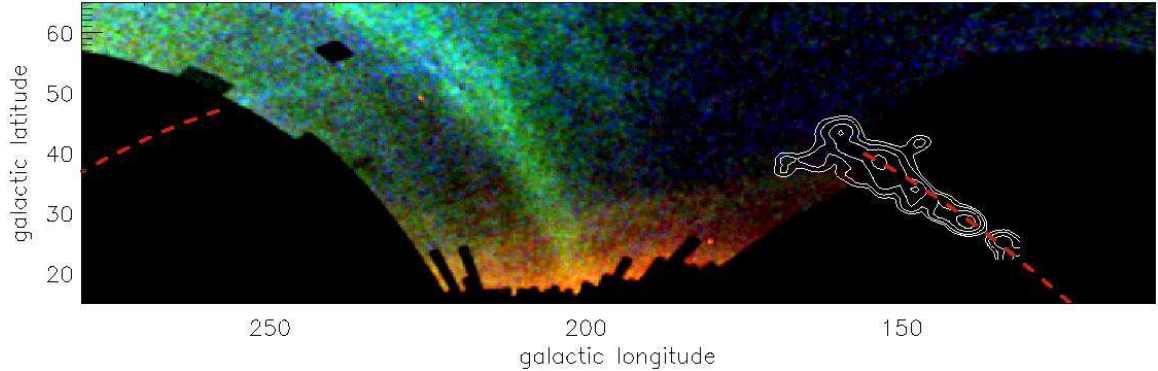}
\caption{\label{fig:complexa} The Orphan Stream in Galactic
coordinates ($\ell,b$). Superposed on the map are the HI column
densities of the Complex A association of High Velocity Clouds from
Wakker (2001). Shown in red is a Galactocentric great circle fit
assuming a orbit radius of $\sim 25$ kpc. }
\end{center}
\end{figure*}
\begin{figure}[t]
\begin{center}
\includegraphics[height=8cm]{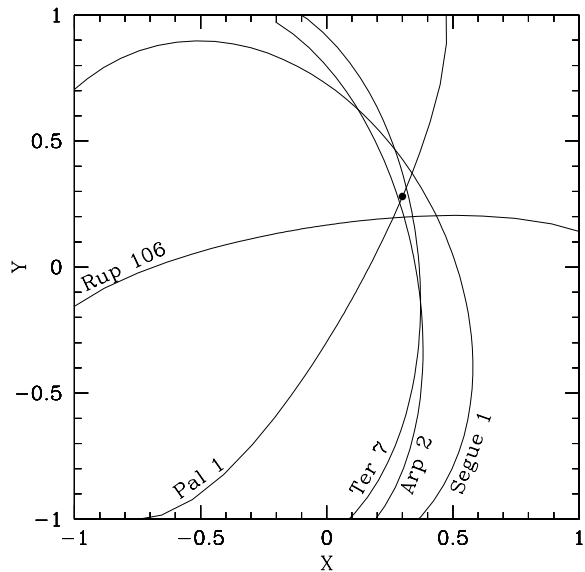}
\caption{\label{fig:wyn} Polar paths of objects passing close to
the pole of the Orphan Stream (marked by a filled circle).}
\end{center}
\end{figure}


\section{The Great Circle of the Orphan Stream}

As pointed out by~\citet{Ly95}, tidal streams lie on great circles
when viewed from the Galactic Center.  The pole of the best fitting
great circle is at $\ellg \approx 42^\circ, \bg \approx 55^\circ$,
where ($\ellg, \bg$) are Galactic coordinates centered on the Galactic
Center.  Figure~\ref{fig:complexa} shows the stream in Galactic
coordinates ($\ell,b$), together with the best fitting great
circle. Superposed on the figures are the contours of HI column
density of an association of High Velocity Clouds (HVCs) known as
Complex A, taken from \cite{Wa01}. This is a stream of HI enshrouding
seven clouds (A0 to AVI) and stretching $\sim 30^\circ$ on the
sky. The arc of neutral gas in Complex A runs along the same great
circle as the optical stream.  Complex A has a distance bracket 4.0 to
10.1 kpc, based on the presence or absence of absorption lines in the
spectra of the stars AD UMa and PG 0859+593~\citep{Wa96,Wo99}.
Although Complex A is closer than the optical stream, it may still be
associated and simply lie on a different wrap of the same orbit. The
velocities of the gas clouds in Complex A range from $-140$ to $-190
\;\kms$. 

Very close to the great circle of the Orphan Stream and behind Complex
A is the recently discovered, disrupting, dwarf spheroidal galaxy,
Ursa Major II, or UMa II~\citep[][ see also Grillmair (2006) who noted
it as a stellar overdensity]{Zu06}. This object lies at Galactic
coordinates $\ell = 152.5^\circ$, $b= 37.4^\circ$. Its color-magnitude
diagram exhibits a well-defined main sequence turn-off, from which its
distance is estimated to be $\sim 30$ kpc, comparable to the distance
of the Orphan Stream as it fades from view. Its radial velocity has
not yet been measured.

\citet{Ly95} developed a method to identify possible globular clusters
associated with a stream. Every possible pole of an object lies at
right angles to its position vector (reckoned from the Galactic
Center). The possible poles sweep out a great circle. Objects that can
lie on the same orbit are identified as intersections in the paths of
the poles of great circles. In practice, it is useful to plot the
polar paths in the coordinates $X = \sqrt{ 1 -\sin \bg}\cos \ellg$ and
$Y = \sqrt{ 1 -\sin \bg}\sin \ellg$ (Lambert's zenithal equal area
projection).  Using the online table of globular cluster data provided
by Harris (1996), we show the polar paths of some objects possibly
associated with the Orphan Stream in Figure~\ref{fig:wyn}.  The pole
of the Orphan Stream lies at $(X \approx 0.31, Y \approx 0.28)$ and is
marked as the filled circle.  It is close to the intersections of
Ruprecht 106, Palomar 1, Arp 2, Terzan 7 and the recently discovered
Segue 1~\citep{Be06c}. Pal 1 and Rup 106 have often been noted as
peculiar.  Both are young halo globular clusters. From isochrone
fitting, \citet{Ro98} estimated that Pal 1 has an age of between 6.3
and 8 Gyr and a metallicity [Fe/H] $\approx -0.6 \pm 0.2$.  Rup 106 is
also younger than typical halo globular clusters, by about 3 to 5 Gyr,
and is very metal poor~\citep{Fr97}. \citet{Pr05} noted that its
$\alpha$ element ratios are significantly lower than Galactic field
stars of similar metallicity. The anomalous properties of Rup 106 had
earlier led \citet{Li92} to propose that it had been accreted from the
Large Magellanic Cloud. Although this is probably not the case, the
idea that the young halo globular clusters may have been accreted from
elsewhere -- possibly from now defunct dwarf galaxies -- has occurred
to a number of investigators~\citep[e.g.,][]{Ly95,Be00,Pr05}.  Terzan
7 and Arp 2 are also young halo globular clusters~\citep{Bu94}, but
their association with the Orphan Stream seems more speculative.  Both
have already been claimed as part of the Sagittarius stream on the
basis of distance, kinematics and chemical composition~\citep[see
e.g.,][]{Sb05}.

\section{Conclusions and Summary}

The Orphan Stream is a $\sim 50^\circ$ arc of stars that was detected
in the Field of Streams by \citet{Be06b}. It was also discovered in
public SDSS data by \citet{Gr06b}, who applied a matched filter
technique to build a color-magnitude diagram and used this to estimate
the average heliocentric distance. His analysis is based on a good
match between the stellar population of the Orphan Stream and the
globular cluster M13.  In this paper, we have presented and analyzed
new observational data on the Orphan Stream, providing continuous
coverage of the Stream through the area where it crosses the
Sagittarius stream.

We carried out a detailed analysis of the available photometric data
and used it to study the stellar populations in the Stream. Both
theoretical isochrones and observational data on three globular
clusters -- M92, M13 and M71 -- were used to build CMD masks. There is
a degeneracy between age, metallicity and distance which cannot be
broken with the existing photometric data. However, there is a strong
indication that the stellar content of the Stream is old and
metal-poor -- similar to, but not identical to, the globular clusters
M92 and M13. A search for blue horizontal branch population was
carried out. Although this did not yield a positive detection,
nonetheless there appears to be a possible blue straggler population
that is associated with the Stream.

We presented evidence for the detection of a distance gradient along
the Stream. The low declination fields are at a heliocentric distance
of $20^{+7}_{-5}$ kpc. At higher declinations, the Stream moves
farther away from us. The last photometric detection of the Stream is
at $32_{-12}^{+15}$ kpc. This is close to the estimated distance of
the newly-discovered dwarf spheroidal galaxy UMa~II, suggesting that
the Orphan Stream may be physically associated with it. \citet{Fe06}
have recently carried out N body simulations of the disruption of
UMa~II and show that its tidal tails match the observational data
available on the Orphan Stream.

Kinematic data can play a crucial role in understanding the nature
of the Orphan Stream. Accordingly, we searched the spectroscopic
database associated with SDSS DR5. This provides radial velocities for
only about $\sim 2 \%$ of all the stars in DR5. Even though there are
very few candidate Orphan Stream stars with spectroscopic data, we
have detected a tentative velocity signal in two fields. At the
celestial equator, the stream is moving towards us at $\sim 40$
kms$^{-1}$. At high declinations, it is moving away from us at $\sim
100$ kms$^{-1}$.

The Orphan Stream lies on the same great circle as Complex A, a linear
association of High Velocity Clouds, as well as a number of globular
clusters, including Ruprecht 106 and Palomar 1. The recently
discovered extended globular cluster Segue 1 is also very close in
position and distance.  All this is consistent with a picture in which
a satellite galaxy merged with the Milky Way long ago. In this
scenario, the Orphan Stream, UMa II, and the young halo globular
clusters were torn off as tidal debris during the merging. Complex A
could be neutral gas that was stripped from a gas-rich dwarf irregular
progenitor, perhaps at a disk-crossing. Alternatively, a large galaxy
can shock and compress ionised gas in the halo, which can then
cool~\citep{Ka66}, leaving a trail of neutral gas in its wake. If so,
then the progenitor must have been massive, with a mass well in excess
of the total mass of Complex A, which is $\sim 10^5 \msun$.

\acknowledgments We particularly wish to thank Bart Wakker and James
Clem for sending us data on Complex A and M92 respectively. Funding
for the SDSS and SDSS-II has been provided by the Alfred P.  Sloan
Foundation, the Participating Institutions, the National Science
Foundation, the U.S. Department of Energy, the National Aeronautics
and Space Administration, the Japanese Monbukagakusho, the Max Planck
Society, and the Higher Education Funding Council for England. The
SDSS Web Site is http://www.sdss.org/.

The SDSS is managed by the Astrophysical Research Consortium for the
Participating Institutions. The Participating Institutions are the
American Museum of Natural History, Astrophysical Institute Potsdam,
University of Basel, Cambridge University, Case Western Reserve
University, University of Chicago, Drexel University, Fermilab, the
Institute for Advanced Study, the Japan Participation Group, Johns
Hopkins University, the Joint Institute for Nuclear Astrophysics, the
Kavli Institute for Particle Astrophysics and Cosmology, the Korean
Scientist Group, the Chinese Academy of Sciences (LAMOST), Los Alamos
National Laboratory, the Max-Planck-Institute for Astronomy (MPIA), the
Max-Planck-Institute for Astrophysics (MPA), New Mexico State
University, Ohio State University, University of Pittsburgh,
University of Portsmouth, Princeton University, the United States
Naval Observatory, and the University of Washington.

\end{document}